\documentclass[prb,reprint,twocolumn,preprintnumbers,amsmath,amssymb,showpacs,footinbib,superscriptaddress]{revtex4-1}

\usepackage[titletoc,toc,title]{appendix}
\usepackage{braket}
\usepackage[final]{feynmp}
\usepackage{ifpdf}
\usepackage{comment}
\usepackage{amsmath}
\usepackage{mathrsfs}
\usepackage{color}
\usepackage{bm}
\usepackage[font=small]{subcaption}
\usepackage[font=small]{caption}

\graphicspath{{./Figures/}}

\makeatletter
\def\endfmffile{%
	\fmfcmd{\p@rcent\space the end.^^J%
			end.^^J%
			endinput;}%
	\if@fmfio
		\immediate\closeout\@outfmf
	\fi
	\ifnum\pdfshellescape=\@ne
		\immediate\write18{mpost \thefmffile}%
	\fi}
\makeatother

\usepackage[demo]{graphicx}
\usepackage{dcolumn}
\usepackage{bm}
\usepackage{hyperref}

\newcommand{\beq}{\begin{equation}}
\newcommand{\eeq}{\end{equation}}

\def\14{{1\over4}}
\def\12{{1 \over 2}}

\def\8{\infty}
\def\oh{\frac{1}{2}}

\def\d{\partial}

\def\undertext#1{\vtop{\hbox{#1}\kern 1pt \hrule}}

\def\be{\begin{equation}}
\def\ee{\end{equation}}
\def\bea{\begin{eqnarray} & &}
\def\eea{\end{eqnarray}}

\def\rf#1{(\ref{#1})}

\def\rf#1{(\ref{#1})}

\def\rfs#1{Eq.~\rf{#1}}

\begin{document}


\title{Theory of avoided criticality in quantum motion in a random potential in high dimensions}

\author{Victor Gurarie}
\affiliation{Department of Physics and Center for Theory of Quantum Matter, University of Colorado, Boulder, Colorado 80309, USA}

\begin{abstract}
The density of states of a three dimensional Dirac equation  with a random potential as well as in other problems of quantum motion in a random potential placed in sufficiently high spatial dimensionality
appears to be singular at a certain critical disorder strength. This was  seen numerically in a variety of studies as well as supported by detailed renormalization group calculations. At the same time it was suggested by a number of arguments accompanied by detailed numerical simulations that this singularity is rounded off by the rare region fluctuations of random potential, and that tuning the disorder past its critical value is not a genuine phase transition but rather a crossover. Here we develop an analytic theory which explains how rare region effects indeed lead to rounding off of the singularity and to the crossover replacing the transition. In particular, this theory relates the
maximum correlation length at the crossover to the disorder induced density of states. 
\end{abstract}

\pacs{Valid PACS appear here}
\maketitle

Thirty years ago E. Fradkin in Ref.~[\onlinecite{Fradkin2}] made a remarkable observation that the density of states in a three dimensional Dirac equation with random scalar potential appears to be a singular function  of energy (a power law of its absolute value) if the disorder strength is tuned to a certain critical value. This observation went against the accepted wisdom that the disorder averaged density of states was typically a smooth function of disorder \cite{Wegner1981,Efetov:book}. At least one example of the models with singular density of states was nonetheless already known in the literature by then (the one-dimensional tight binding model with random hoppings studied in Ref.~[\onlinecite{Eggarter1978}] which has the so-called Dyson singularity in its density of states at zero energy). It was subsequently understood that this and other similar models  have 
extra symmetries systematically studied by  Altland and Zirnbauer \cite{Altland1997}, resulting in their now famous classification table. The three dimensional Dirac equation studied in Ref.~[\onlinecite{Fradkin2}] does not appear to fall under any of the non-trivial entries of this table, so the potential singularity of its density of states if present is unrelated to the Altland-Zirnbauer classification table. 

Interest to the three dimensional Dirac equation with random scalar potential was revived recently due to the theoretical prediction \cite{Wan:WeylProp} and observation 
\cite{ZHasan:TaAs,ZHasan:TaAs2,Weng:PhotCrystWSM,Cava2014,Yazdani2014} of the  semi-metals with the band structures well described by  three dimensional Weyl or Dirac equations. Theoretical \cite{Goswami:TIRG} (in the framework of renormalization group (RG)) and numerical \cite{Herbut} studies seemingly confirmed the presence of the singularity first predicted by Fradkin. A number of studies followed, for example Refs.~[\onlinecite{Brouwer:exponents,Brouwer:WSMcond,RoyDasSarma,Pixley:ExactZ,Syzranov:Weyl,Syzranov:exp}], exploring the critical properties of this singularity and its effects on transport in the Dirac equation and other phenomena.

Subsequently Ref.~[\onlinecite{Syzranov:unconv}] generalized the arguments of Ref.~\onlinecite{Fradkin2}] to show that any quantum problem with random scalar potential, if placed in space of high enough dimension $d>d_c$, develops singular density of states at a certain critical disorder strength. In particular, for the conventional Schr\"odinger equation with quenched disorder, this occurs with $d_c=4$, for Dirac equation $d_c=2$, and models can be developed which have arbitrary $d_c$, even with $d_c<1$ if desired \cite{Garttner:longrange}.  One striking prediction of the Ref.~[\onlinecite{Syzranov:unconv}] concerned the nature of the Anderson transition \cite{Efetov:book} in the disordered Scr\"odinger equation at $d>4$ \cite{Slevin:high}, where if disorder is tuned to its critical value, the divergence of the localization length as energy approaches the mobility edge was predicted
to be controlled by an exponent distinct from the conventional $d$-dependent exponent  of the Anderson transition. The latter was predicted to be at work only if disorder strength was above its critical value. Finally, Ref.~[\onlinecite{Syzranov:mult}] showed, in part by adopting the arguments of Ref.~[\onlinecite{Foster2012}] to higher spatial dimensions, that at disorder and energy tuned to its critical value, the wave functions of these problems are multifractal. 

However, it was noted in an number of publications that all these models lack an order parameter which would label the different phases separated by the critical point discussed above (often referred in the literature as ``ballistic" and ``diffusive" phases, occurring at weak and strong disorder respectively). Specifically, let us take the three dimensional Weyl problem governed by the Hamiltonian
\begin{equation} \label{eq:Weyl} H = - i v \sum_{j=1}^3 \sigma^j \d_j + V({\bf r}), \end{equation}
where $\sigma^j$ are Pauli matrices, and $V({\bf r})$ is the random potential, and look at its disorder averaged density of states $\rho(E)$, defined as always by
\be \rho(E) =\frac 1 {L^d} \sum_m \left< \delta(E-E_m) \right>,
\ee
where $L$ is the size of the system, $E_m$ are exact energy levels, and the brackets denote averaging over the realizations of the random potential. 

Ref.~[\onlinecite{Fradkin2}] argued that $\rho(0)=0$ if the disorder strength is below its critical value and $\rho(0)>0$ if it is above that value, in case when the random potential averages to zero $\left< V({\bf r}) \right>=0$. This seems to be supported by some of the available numerical studies of this model \cite{Herbut}. Thus we could take $\rho(0)$ as the  order parameter of the transition potentially occurring in this model. However, it is clear that $\rho(0)$ cannot be exactly zero at any disorder strength. Indeed, a fluctuation of disorder which is constant over some (large) region of space, which one can term a rare region, shifts the energy of the eigenstates of the \rfs{eq:Weyl} resulting in some of them shifted towards zero energy, rendering $\rho(0)$ nonzero  \cite{Nandkishore:rare}, an effect neglected in Ref.~[\onlinecite{Fradkin2}]. This invalidates using $\rho(0)$ as the order parameter. In the absence of other suitable candidates, one may conclude that there could not be a transition in this problem with a critical density of states. Yet this argument does not unambiguously rule out a genuine singularity in the density of states at a critical value of disorder. Despite this, subsequent numerical studies going beyond the earlier work of Ref.~[\onlinecite{Herbut}], starting from the work by Ref.~[\onlinecite{Pixley2016}], together with some analytic arguments were able to resolve the absence of singularity in the density of states \cite{Pixley2016,Pixley2016a,Radzihovsky2017,Ostrovsky2017} and argue in favor of the crossover replacing the transition in the model \rfs{eq:Weyl}. 

Note that even if the transition is indeed absent and replaced by a crossover, this does not invalidate the prior studies of criticality and multifractality in these models. It just implies that unlike in genuine critical points, in these problems as the disorder strength is tuned to what should be its critical value,  the correlation length saturates at some maximum value $\ell$ instead of diverging to infinity. All the critical scaling and multifractality will still be observed at scales below that saturation scale, explaining why most numerical studies failed to distinguish the crossover in these models from genuine critical points. In particular, the Anderson transition in the model \rfs{eq:schr} at $d>d_c=4$ is still governed by the Anderson critical exponent, however at disorder strength close to what would have been its critical value, a crossover  regime at $E$ below the mobility edge should exist where the localization length is governed by the exponent discussed in Ref.~[\onlinecite{Syzranov:unconv}], turning into the conventional Anderson exponent as $E$ approaches
the mobility edge\footnote{The correlation length, being the length of the spatial decay of the single particle Green's function, should be identified with the mean free path of a particle in disordered environment. From this point of view, it is indeed natural that it does not diverge to infinity; otherwise one would be forced to conclude that the mean free path could be infinite at a finite disorder strength.}.

Here we would like to present quantitative arguments showing that in these problems there cannot be a genuine singularity in the density of states and develop an analytic  framework to calculate the
saturation scale which controls these crossover phenomena.


Let us examine the  Schr\"odinger equation with random potential at $d>d_c=4$, defined by the usual Hamiltonian
\be \label{eq:schr} H = -\frac{1}{2m} \Delta + V({\bf r}).
\ee
Ref.~[\onlinecite{Syzranov:unconv}] argued that since, just like in the Dirac problem at $d=3$ studied in Ref.~[\onlinecite{Fradkin2}] and in subsequent studies, the weak disorder here is irrelevant in the RG sense, while strong disorder is relevant, there should be a critical disorder strength separating these two phases. At this critical disorder strength, $\rho(E)$ for this problem should have a singularity at $E=E_c$ (unlike in the Dirac problem considered above, $E_c$ may be different from zero). This can be encoded by the following RG equation satisfied by the dimensionless disorder strength $\gamma=\lambda \xi^{4-d}$ defined by the correlator
\be \label{eq:dc} \left< V({\bf r}) V({\bf r}') \right> = 2 \lambda \, \delta({\bf r}- {\bf r}' ),
\ee which can be shown to be \cite{Syzranov:unconv}
\be \label{eq:rg}\dot \gamma = -(d-4) \gamma + C \gamma^2  + \dots,
\ee where $C>0$ is some positive constant (as always, $\xi$ is the ultraviolet length scale defined  precisely below). This equation seemingly shows that at $d>4$, weak $\gamma$ is irrelevant and flows to the disorder-free $\gamma=0$ fixed point, while there exist a value $\gamma=\gamma_c$ which nullifies the right hand side of \rfs{eq:rg} and which correspond to the critical point in the renormalization group (as always, \rfs{eq:rg} is perturbative in $\gamma$, so $\gamma_c$ can be extracted from it only if $d-4$ is small and $\gamma_c$ is small enough for the perturbation theory to work, but usually $\gamma_c$
which nullifies the right-hand-side of \rfs{eq:rg} is presumed to exist at any $d>4$). It is at this critical point $\gamma=\gamma_c$, corresponding to $\lambda=\lambda_c=
\gamma_c \xi^{d-4}$, that we expect the density of states to be a critical function of energy.

These arguments immediately lead to a striking prediction that disagrees with the available literature. It is well known the disorder-averaged Green's function of the problem defined 
by \rf{eq:schr} and \rf{eq:dc} is equivalent to the self-attracting random walks where $\lambda$ controls the strength of the attraction \cite{deGennes1972,Emery1975} (while $\lambda<0$ corresponds
to the more familiar case of self-repulsive random walks). 
It is tempting to conclude from \rfs{eq:rg} that the self-attracting random walks
at $d>4$ 
are equivalent to Brownian motion for the attraction below critical, and are critical at the critical value of the attraction which nullifies the right hand side of \rfs{eq:rg}. However, existing studies of
self-attractive random walks at $d>4$ showed that those are never equivalent to Brownian motion, and do not have any special critical value of the strength of the attraction \cite{Oono1975,Oono1976,Slade1995}. 
By extension, this hints at the absence of criticality in the random Schr\"odinger equation at $d>4$.

While adopting the arguments of Refs.~[\onlinecite{Oono1975,Oono1976,Slade1995}] to the motion in random potential at $d>4$ is possible to explain the absence of criticality
in \rfs{eq:schr}, the  drawback of this approach is that it is not easy to apply it to other problems such as 
the Weyl equation in three dimensions with disorder defined by \rfs{eq:Weyl} or the one dimensional systems discussed in Ref.~[\onlinecite{Garttner:longrange}]. 

We would therefore like to present another argument based on adopting the method of instantons in the replica approach to the problem of calculating the density of states. It has long been known that these instantons is what produces the so-called Lifshitz tail \cite{Lifshitz:tail} density of states where there would have been none had the disorder been absent. We would like to show that these instantons, when properly resummed, result in the absence of any singularity in the density of states of these problems, thus unifying the absence of vanishing density of states and the absence of criticality as anticipated in Refs.~[\onlinecite{Nandkishore:rare,Radzihovsky2017}]. 

While this argument should work equally well for \rfs{eq:schr} or \rfs{eq:Weyl}, we will demonstrate it in case of the random Schr\"odinger equation for simplicity (leaving the task of adopting it for \rfs{eq:Weyl} to future work). Taking Eqs.~\rf{eq:schr} and \rf{eq:dc} we write down the replica theory which produces the disorder averaged retarded Green's function
\be \label{eq:part} Z = \int {\cal D} \phi \, e^{\frac i 2 \int  d^d x \sum_n \phi_n \left( E + i0 + \Delta \right) \phi_n - \frac  \lambda 4 \left(  \sum_n \phi_n^2 \right)^2}
\ee
(from now on we set $m=1/2$ since it can always be put back into every equation simply by dimensional analysis). 
The density of states for $E\le 0$, where in the absence of disorder potential the density of states is zero, can be computed using the instanton or saddle point method as first discussed in Ref.~[\onlinecite{ZittartzLanger}]. Extremizing  the action in the exponential and following Ref.~[\onlinecite{Cardy1978}], we find 
\be \label{eq:ns} \Delta f = \left| E \right| f - \lambda f^3,
\ee
where $\phi_n = v_n f({\bf r})(1-i)/\sqrt{2}$, and $v_n$ being a unit vector in replica space. Given a solution of this equation, the action in the exponential of \rfs{eq:part} evaluates to 
\be \label{eq:actcalc} S = \frac{\lambda}{4} \int d^dx \, f^4,
\ee and produces the contribution to the density of states
\be \rho \sim e^{-S}.
\ee

If $d<d_c=4$, as explained in Ref.~[\onlinecite{Cardy1978}], a theorem proven in Ref.~[\onlinecite{Coleman1978}] guarantees that there is a finite action solution to \rfs{eq:ns} of the form
\be \label{eq:sca} f = \sqrt{ \left| E \right| /\lambda} \, F \left( r \sqrt{\left| E \right|} \right),
\ee
where $F$ is a dimensionless function finite at the origin and quickly decaying at infinity, producing the finite action and the density of states
\be \label{eq:lif} \rho \sim \exp\left( -{\rm const}  \left| E \right|^{2-d/2}/\lambda \right),
\ee
with the numerical constant which can be found if $F$ is known (here and below, the proportionality symbol implies equality up to an overall numerical coefficient).
This  result is well known in the theory of Lifshitz tail states. 

We now focus on the regime $d \ge d_c=4$, where the theorem of Ref.~[\onlinecite{Coleman1978}] no longer applies. We investigated \rfs{eq:ns} numerically to find that most likely all solutions of the form \rfs{eq:sca} produce an infinite action. The insight
of Ref.~[\onlinecite{Nandkishore:rare}], which can be taken from their analysis of the corresponding problem arising in the context of \rfs{eq:Weyl}, is that we should instead broaden the correlation function of disorder
in \rfs{eq:dc} and consider disorder correlated at finite length $\xi$
\be \label{eq:kernel}
\left< V({\bf r}) V({\bf r}') \right> =2 \lambda \xi^{-d} \, K(\left|{\bf r}- {\bf r}'\right|/\xi ),
\ee
where $K$ is a dimensionless function of its argument decaying to zero at large arguments such that
\be \int d^d r \, K(r) = 1,
\ee so that $\xi \rightarrow 0$ limit of $K$ is just the original delta function. The equation \rfs{eq:ns} now becomes
\be \label{eq:nsi}  \Delta f({\bf r}) = \left| E \right| f({\bf r}) - \frac{\lambda f({\bf r})}{\xi^d} \int d^d r' \, K\left(\left| {\bf r}- {\bf r}'\right|/\xi \right) f^2({\bf r}').
\ee
Let us study \rfs{eq:nsi} at $E=0$. There exists a solution $f_{\rm inst}$ of this equation which, at large distances, coincides with the solution of the Laplace equation in $d$-dimensional space
\be \label{eq:coul} f_{\rm inst} (r) = \frac{A}{r^{d-2}}, \ r \gg \xi.
\ee
Importantly, at $r \gg \xi$, the right hand side of \rfs{eq:nsi} goes as $1/r^{3(d-2)}$ while the terms on the left hand side go as $1/r^d$. At $3(d-2)>d$ or $d>3$ (which is the regime we are interested in) the right hand side can be neglected, making \rfs{eq:coul} a solution. At $r \sim \xi$, we have a consistency condition which follows from matching the left and the right hand sides of \rfs{eq:nsi}, giving 
\be \frac{A}{\xi^d} \sim \frac{\lambda A^3}{\xi^{3(d-2)}}, \, A \sim \xi^{d - 3}/\sqrt{\lambda}.
\ee
At distances $r \lesssim \xi$, the right hand side of \rfs{eq:nsi} cannot be neglected. Its effect is to regularize the solution and make it nonsingular at $r=0$. This results in the solution $f_{\rm inst}({\bf r})$ to \rfs{eq:nsi} at $E=0$ such that
\be \label{eq:finst} f_{\rm inst} (r) = A_0 \frac{\xi^{d-3}}{\sqrt{\lambda}} \frac 1 { r^{d-2}}, \  r \gg \xi,
\ee where $A_0$ is a numerical coefficient and $f(r)$ finite at $r \rightarrow 0$
The action of this solution is the generalization of \rfs{eq:actcalc} onto the case of finite disorder correlation \rfs{eq:kernel}, which is
\be \label{eq:actionkernel} S = \frac{\lambda}{4} \int \frac{{d^d r} {d^d r'}}{\xi^d} K\left( \left| {\bf r} - {\bf r}' \right| /\xi \right) f^2({\bf r}) f^2\left({\bf r}'\right).
\ee
Evaluating the action for $f=f_{\rm inst}$, we observe that $K$ can still be approximated as a delta-function at $r \gg \xi$, as emphasized in Ref.~[\onlinecite{Nandkishore:rare}], to give an estimate
\be \label{eq:singleact} S \sim A_0^4 \frac{\xi^{4d - 12}}{\lambda} \int_{\xi}^\infty \frac{r^{d-1} dr }{r^{4d-8}} \sim \frac{\xi^{d-4}}{\lambda}.
\ee
This produces the density of states at $E=0$ as 
\be \label{eq:doshighzero} \rho \sim \exp(-{\rm const} \, \xi^{d-4}/\lambda),\ee in agreement with Refs.~[\onlinecite{Suslov1994}] and [\onlinecite{Syzranov:unconv}] (for uncorrelated potential, $\xi$ should be taken to
be equal to lattice constant). Note that $E \not = 0$ results in a modified $f_{\rm inst}$  
 at $r \gg |E|^{-1/2}$ where it now decays exponentially, which as a first approximation does not change \rfs{eq:singleact}, also in
agreement with Refs.~[\onlinecite{Suslov1994}] and [\onlinecite{Syzranov:unconv}].

We are now in position to address the central issue of this paper: how the instanton solutions of this type can make the Green's function  to decay exponentially in space. A single instanton solution produces an additive contribution to $Z$ and to the Green's function, thus resulting in the density of states which is the sum of the instanton (smooth) and the perturbative (singular) parts. However, we can generalize the solution \rfs{eq:coul} to a gas of instantons of the form
\be \label{eq:gas} f({\bf r}) = \sum_{n=1}^N \zeta_n f_{\rm inst } \left({\bf r}- {\bf r}_n \right).
\ee
where $\zeta_n = \pm 1$ and ${\bf r}_n$ are positions of instantons, and $N$ is their number. Importantly, as long as the instantons are apart by more than $\xi$, \rfs{eq:gas} approximately solves \rfs{eq:nsi}. Substituting this into \rfs{eq:actcalc} we note that terms of the form $1/\left| {\bf r} -{\bf r}_n \right|^{4(d-2)}$ produce the sum of the actions \rfs{eq:singleact}, one per instanton, representing the sum of  their ``core energies". The next term comes from the cross terms of the form 
\be \label{eq:instint}\frac 1  \lambda \int  \frac{  \zeta_n \zeta_m \xi^{4d - 12}d^d r }{ \left| {\bf r} -{\bf r}_n \right|^{3(d-2)} \left| {\bf r} -{\bf r}_m \right|^{d-2}} \sim \frac{\zeta_n \zeta_m \xi^{2d-6}}{\lambda \left| {\bf r}_m -{\bf r}_n \right|^{d-2}}.
\ee
To evaluate this integral we have to remember that the functions $1/r^{d-2}$ are all understood as being regularized at $r \lesssim \xi$, and the integrals can be estimated by integrating over ${\bf r}$ in the vicinity of ${\bf r}_n$. The rest of the terms produced by \rfs{eq:actcalc} with \rfs{eq:gas} substituted will have weaker divergencies as ${\bf r}$ approaches ${\bf r}_n$, leading to terms smaller that the one above at small $\xi$. More
details related to the derivation of \rfs{eq:instint} are given in the Appedix \ref{sec:A}, see \rfs{eq:actionexplained} to \rf{eq:twobo}. 

The conclusion is that these instantons interact via two body Coulomb-like interaction $1/r^{d-2}$ given by \rfs{eq:instint}, with three and four body terms suppressed as powers of $\xi$, leading to a Coulomb gas of instantons. These have been studied at length in the literature, in particular in dimensionalities\cite{Polyakov1977} $d>2$. 
The standard 
approach to the Coulomb gas at $d>2$ described for example in Ref.~[\onlinecite{Polyakov:book}] can now almost literally be adapted to our problem. This gas of instantons can be recast in the form of an effective field theory with the action
\be \label{eq:instanton} S =  \int d^d x \, \left[ \frac{D \lambda \xi^{6-2d}}{2}  \left( \nabla \chi \right)^2 -i \mu  \xi^{-d} \cos(\chi)  \right],
\ee
where $D$ is a dimensionless factor and a real dimensionless $\mu$ can be related to the density of states (the details of the derivation of Eqs.~\rf{eq:instint} and \rf{eq:instanton}, while mostly standard, are given in the Appendix \ref{sec:A}). 

$\cos(\chi)$ is a relevant perturbation, so it can be expanded $\cos(\chi) \approx 1 - \chi^2/2$ leading to \rfs{eq:instanton} describing exponentially decaying (in space) Green's function  (see Appendix~\ref{sec:A} and \rf{eq:greenfull} for 
the detailed derivation)
\be \label{eq:gmain} G(p) \sim \left[    p^2 + i \mu \xi^{d-6}/(D \lambda) \right]^{-1}.
\ee
with the correlation length
\be \ell \sim  \xi^{3-d/2} \sqrt{\lambda/\mu}
\ee (for the derivation of \rfs{eq:gmain} including the disorder-dependent prefactor omitted here see Appendix \ref{sec:A} leading to \rfs{eq:greenfull}).
We note that the density of states at zero energy can now be calculated simply as
\be \label{eq:rho0} \rho(0) =\frac 1 \pi {\rm Im} \, \int \frac{d^dp}{(2\pi)^d} \frac{1}{ p^2+i \mu \xi^{d-6}/(D\lambda) } \sim \frac{\mu}{\lambda \xi^2} ,
\ee
Importantly, as $d>d_c$, this is an ultraviolently divergent integral so to estimate it we just need to expand it in powers of $\mu$ and keep the leading term (as always the divergence is cut off 
since $p$ in Eq.~\rf{eq:rho0} cannot exceed  $1/\xi$). 
We already calculated the density of states in \rfs{eq:doshighzero}. The new relation \rfs{eq:rho0} by connecting it to $\mu$ allows us to relate it to the correlation length. 
We thus find the correlation length to be
\be \ell \sim \frac{1} { \xi^{\frac d 2 - 2}\sqrt{\rho(0)}},
\ee
giving us a relationship between the density of states at zero energy (which would have been zero in the absence of disorder and is only nonzero due to rare fluctuations of disorder) and the correlation length, the 
length at which the divergent length scales saturate in the vicinity of $\lambda=\lambda_c$. 
Technically speaking, the line of arguments presented here applies at very small $\lambda$. In practice however $\rho(0)$ is known to be small (numerically, for example), and we can rely on this to justify
the instanton approximation even when $\lambda \simeq \lambda_c$.  

This derivation can easily be generalized to the problems of particles whose disorder-free spectrum is $ E = p^\alpha$ with the space dimensionality $d>2\alpha$, where $\alpha=2$ for \rfs{eq:schr}, $\alpha=1$ for \rfs{eq:Weyl} and to models where $\alpha$ can be freely tuned described in Ref.~[onlinecite{Garttner:longrange}].  \rfs{eq:rg} gets modified to \be \dot \gamma = - (d-2 \alpha) \gamma + C \gamma^2+ \dots,\ee  with the critical point $\gamma_c$ occurring at $d>2\alpha$ ($\gamma= \lambda \xi^{2\alpha-d}$). 

Repeating the arguments above we find that the single instanton solution goes as \be f_{\rm inst}(r) \sim \xi^{d-3\alpha/2}/(r^{d-\alpha} \sqrt{\lambda}),\ee  leading to the density of states 
at zero energy
\be \rho \sim \exp(-{\rm const} \, \xi^{d-2\alpha}/\lambda)\ee (this matches the density of states of \rfs{eq:Weyl} derived in Ref.~[\onlinecite{Nandkishore:rare}] if $d=3$, $\alpha=1$ is substituted, and of course matches \rfs{eq:doshighzero} if $\alpha=2$). A gas of instantons will interact  with the pair-wise potential $\xi^{2d-3\alpha}/( \lambda \, r^{d-\alpha})$. This produces the effective field theory with the Green's function
\be \label{eq:gralpj} G(p) \sim\frac{ D \lambda \xi^{3 \alpha-2d}}{D \lambda \xi^{3 \alpha-2d} p^\alpha + i \mu \xi^{-d} },
\ee
and the correlation length \be \ell \sim \xi^{3-d/\alpha} (\lambda/\mu)^{1/\alpha}.\ee  Importantly, the density of states is still proportional to $\mu$, 
\be \rho(0) \sim \frac{\mu}{\lambda \xi^{\alpha}},\ee therefore \be \label{eq:dosalll} \ell \sim \frac {1} {\xi^{\frac{d}{\alpha}-2} \rho(0)^{1/\alpha}}.\ee In particular, in the 
important application of this formalism to the Weyl problem defined by \rfs{eq:Weyl} and \rfs{eq:dc}, \be \ell \sim \frac{1}{\xi \rho(0)}. \ee
The details of this derivation can be found in Appendix \ref{sec:B}. 

With the help of the formalism developed here, it is now possible to study the physical consequences of the avoided criticality in the disordered Weyl problem, which will be the subject of future research.

The author is grateful to L. Radzihovsky, S. Syzranov and R. Nandkishore for many useful discussions concerning this problem and acknowledges support  by the NSF grant DMR-1205303.





\newpage 

\bibliography{instantons}

\begin{thebibliography}{42}%
\makeatletter
\providecommand \@ifxundefined [1]{%
 \@ifx{#1\undefined}
}%
\providecommand \@ifnum [1]{%
 \ifnum #1\expandafter \@firstoftwo
 \else \expandafter \@secondoftwo
 \fi
}%
\providecommand \@ifx [1]{%
 \ifx #1\expandafter \@firstoftwo
 \else \expandafter \@secondoftwo
 \fi
}%
\providecommand \natexlab [1]{#1}%
\providecommand \enquote  [1]{``#1''}%
\providecommand \bibnamefont  [1]{#1}%
\providecommand \bibfnamefont [1]{#1}%
\providecommand \citenamefont [1]{#1}%
\providecommand \href@noop [0]{\@secondoftwo}%
\providecommand \href [0]{\begingroup \@sanitize@url \@href}%
\providecommand \@href[1]{\@@startlink{#1}\@@href}%
\providecommand \@@href[1]{\endgroup#1\@@endlink}%
\providecommand \@sanitize@url [0]{\catcode `\\12\catcode `\$12\catcode
  `\&12\catcode `\#12\catcode `\^12\catcode `\_12\catcode `\%12\relax}%
\providecommand \@@startlink[1]{}%
\providecommand \@@endlink[0]{}%
\providecommand \url  [0]{\begingroup\@sanitize@url \@url }%
\providecommand \@url [1]{\endgroup\@href {#1}{\urlprefix }}%
\providecommand \urlprefix  [0]{URL }%
\providecommand \Eprint [0]{\href }%
\providecommand \doibase [0]{http://dx.doi.org/}%
\providecommand \selectlanguage [0]{\@gobble}%
\providecommand \bibinfo  [0]{\@secondoftwo}%
\providecommand \bibfield  [0]{\@secondoftwo}%
\providecommand \translation [1]{[#1]}%
\providecommand \BibitemOpen [0]{}%
\providecommand \bibitemStop [0]{}%
\providecommand \bibitemNoStop [0]{.\EOS\space}%
\providecommand \EOS [0]{\spacefactor3000\relax}%
\providecommand \BibitemShut  [1]{\csname bibitem#1\endcsname}%
\let\auto@bib@innerbib\@empty
\bibitem [{\citenamefont {Fradkin}(1986)}]{Fradkin2}%
  \BibitemOpen
  \bibfield  {author} {\bibinfo {author} {\bibfnamefont {E.}~\bibnamefont
  {Fradkin}},\ }\href@noop {} {\bibfield  {journal} {\bibinfo  {journal} {Phys.
  Rev. B}\ }\textbf {\bibinfo {volume} {33}},\ \bibinfo {pages} {3257}
  (\bibinfo {year} {1986})}\BibitemShut {NoStop}%
\bibitem [{\citenamefont {Wegner}(1981)}]{Wegner1981}%
  \BibitemOpen
  \bibfield  {author} {\bibinfo {author} {\bibfnamefont {F.}~\bibnamefont
  {Wegner}},\ }\href@noop {} {\bibfield  {journal} {\bibinfo  {journal} {Z
  Physik B}\ }\textbf {\bibinfo {volume} {44}},\ \bibinfo {pages} {9} (\bibinfo
  {year} {1981})}\BibitemShut {NoStop}%
\bibitem [{\citenamefont {Efetov}(1999)}]{Efetov:book}%
  \BibitemOpen
  \bibfield  {author} {\bibinfo {author} {\bibfnamefont {K.~B.}\ \bibnamefont
  {Efetov}},\ }\href@noop {} {\emph {\bibinfo {title} {Supersymetry in Disorder
  and Chaos}}}\ (\bibinfo  {publisher} {Cambridge University Press},\ \bibinfo
  {address} {New York},\ \bibinfo {year} {1999})\BibitemShut {NoStop}%
\bibitem [{\citenamefont {Eggarter}\ and\ \citenamefont
  {Riedinger}(1978)}]{Eggarter1978}%
  \BibitemOpen
  \bibfield  {author} {\bibinfo {author} {\bibfnamefont {T.~P.}\ \bibnamefont
  {Eggarter}}\ and\ \bibinfo {author} {\bibfnamefont {R.}~\bibnamefont
  {Riedinger}},\ }\href@noop {} {\bibfield  {journal} {\bibinfo  {journal}
  {Phys. Rev. B}\ }\textbf {\bibinfo {volume} {18}},\ \bibinfo {pages} {569}
  (\bibinfo {year} {1978})}\BibitemShut {NoStop}%
\bibitem [{\citenamefont {Altland}\ and\ \citenamefont
  {Zirnbauer}(1997)}]{Altland1997}%
  \BibitemOpen
  \bibfield  {author} {\bibinfo {author} {\bibfnamefont {A.}~\bibnamefont
  {Altland}}\ and\ \bibinfo {author} {\bibfnamefont {M.}~\bibnamefont
  {Zirnbauer}},\ }\href@noop {} {\bibfield  {journal} {\bibinfo  {journal}
  {Phys. Rev. B}\ }\textbf {\bibinfo {volume} {55}},\ \bibinfo {pages} {1142}
  (\bibinfo {year} {1997})}\BibitemShut {NoStop}%
\bibitem [{\citenamefont {Wan}\ \emph {et~al.}(2011)\citenamefont {Wan},
  \citenamefont {Turner}, \citenamefont {Vishwanath},\ and\ \citenamefont
  {Savrasov}}]{Wan:WeylProp}%
  \BibitemOpen
  \bibfield  {author} {\bibinfo {author} {\bibfnamefont {X.}~\bibnamefont
  {Wan}}, \bibinfo {author} {\bibfnamefont {A.~M.}\ \bibnamefont {Turner}},
  \bibinfo {author} {\bibfnamefont {A.}~\bibnamefont {Vishwanath}}, \ and\
  \bibinfo {author} {\bibfnamefont {S.~Y.}\ \bibnamefont {Savrasov}},\
  }\href@noop {} {\bibfield  {journal} {\bibinfo  {journal} {Phys. Rev. B}\
  }\textbf {\bibinfo {volume} {83}},\ \bibinfo {pages} {205101} (\bibinfo
  {year} {2011})}\BibitemShut {NoStop}%
\bibitem [{\citenamefont {Huang}\ \emph {et~al.}(2015)\citenamefont {Huang},
  \citenamefont {Xu}, \citenamefont {Belopolski}, \citenamefont {Lee},
  \citenamefont {Chang}, \citenamefont {Wang}, \citenamefont {Alidoust},
  \citenamefont {Bian}, \citenamefont {Neupane}, \citenamefont {Zhang},
  \citenamefont {Jia}, \citenamefont {Bansil}, \citenamefont {Lin},\ and\
  \citenamefont {Hasan}}]{ZHasan:TaAs}%
  \BibitemOpen
  \bibfield  {author} {\bibinfo {author} {\bibfnamefont {S.-M.}\ \bibnamefont
  {Huang}}, \bibinfo {author} {\bibfnamefont {S.-Y.}\ \bibnamefont {Xu}},
  \bibinfo {author} {\bibfnamefont {I.}~\bibnamefont {Belopolski}}, \bibinfo
  {author} {\bibfnamefont {C.-C.}\ \bibnamefont {Lee}}, \bibinfo {author}
  {\bibfnamefont {G.}~\bibnamefont {Chang}}, \bibinfo {author} {\bibfnamefont
  {B.}~\bibnamefont {Wang}}, \bibinfo {author} {\bibfnamefont {N.}~\bibnamefont
  {Alidoust}}, \bibinfo {author} {\bibfnamefont {G.}~\bibnamefont {Bian}},
  \bibinfo {author} {\bibfnamefont {M.}~\bibnamefont {Neupane}}, \bibinfo
  {author} {\bibfnamefont {C.}~\bibnamefont {Zhang}}, \bibinfo {author}
  {\bibfnamefont {S.}~\bibnamefont {Jia}}, \bibinfo {author} {\bibfnamefont
  {A.}~\bibnamefont {Bansil}}, \bibinfo {author} {\bibfnamefont
  {H.}~\bibnamefont {Lin}}, \ and\ \bibinfo {author} {\bibfnamefont {M.~Z.}\
  \bibnamefont {Hasan}},\ }\href@noop {} {\bibfield  {journal} {\bibinfo
  {journal} {Nature Comm.}\ }\textbf {\bibinfo {volume} {6}},\ \bibinfo {pages}
  {7373} (\bibinfo {year} {2015})}\BibitemShut {NoStop}%
\bibitem [{\citenamefont {Xu}\ \emph {et~al.}(2015)\citenamefont {Xu},
  \citenamefont {Belopolski}, \citenamefont {Alidoust}, \citenamefont
  {Neupane}, \citenamefont {Bian}, \citenamefont {Zhang}, \citenamefont
  {Sankar}, \citenamefont {Chang}, \citenamefont {Yuan}, \citenamefont {Lee},
  \citenamefont {Huang}, \citenamefont {Zheng}, \citenamefont {Ma},
  \citenamefont {Sanchez}, \citenamefont {Wang}, \citenamefont {Bansil},
  \citenamefont {Chou}, \citenamefont {Shibayev}, \citenamefont {Lin},
  \citenamefont {Jia},\ and\ \citenamefont {Hasan}}]{ZHasan:TaAs2}%
  \BibitemOpen
  \bibfield  {author} {\bibinfo {author} {\bibfnamefont {S.-Y.}\ \bibnamefont
  {Xu}}, \bibinfo {author} {\bibfnamefont {I.}~\bibnamefont {Belopolski}},
  \bibinfo {author} {\bibfnamefont {N.}~\bibnamefont {Alidoust}}, \bibinfo
  {author} {\bibfnamefont {M.}~\bibnamefont {Neupane}}, \bibinfo {author}
  {\bibfnamefont {G.}~\bibnamefont {Bian}}, \bibinfo {author} {\bibfnamefont
  {C.}~\bibnamefont {Zhang}}, \bibinfo {author} {\bibfnamefont
  {R.}~\bibnamefont {Sankar}}, \bibinfo {author} {\bibfnamefont
  {G.}~\bibnamefont {Chang}}, \bibinfo {author} {\bibfnamefont
  {Z.}~\bibnamefont {Yuan}}, \bibinfo {author} {\bibfnamefont {C.-C.}\
  \bibnamefont {Lee}}, \bibinfo {author} {\bibfnamefont {S.-M.}\ \bibnamefont
  {Huang}}, \bibinfo {author} {\bibfnamefont {H.}~\bibnamefont {Zheng}},
  \bibinfo {author} {\bibfnamefont {J.}~\bibnamefont {Ma}}, \bibinfo {author}
  {\bibfnamefont {D.~S.}\ \bibnamefont {Sanchez}}, \bibinfo {author}
  {\bibfnamefont {B.}~\bibnamefont {Wang}}, \bibinfo {author} {\bibfnamefont
  {A.}~\bibnamefont {Bansil}}, \bibinfo {author} {\bibfnamefont
  {F.}~\bibnamefont {Chou}}, \bibinfo {author} {\bibfnamefont {P.~P.}\
  \bibnamefont {Shibayev}}, \bibinfo {author} {\bibfnamefont {H.}~\bibnamefont
  {Lin}}, \bibinfo {author} {\bibfnamefont {S.}~\bibnamefont {Jia}}, \ and\
  \bibinfo {author} {\bibfnamefont {M.~Z.}\ \bibnamefont {Hasan}},\ }\href@noop
  {} {\bibfield  {journal} {\bibinfo  {journal} {Science}\ }\textbf {\bibinfo
  {volume} {349}},\ \bibinfo {pages} {6248} (\bibinfo {year}
  {2015})}\BibitemShut {NoStop}%
\bibitem [{\citenamefont {Lv}\ \emph {et~al.}(2015)\citenamefont {Lv},
  \citenamefont {Weng}, \citenamefont {Fu}, \citenamefont {Wang}, \citenamefont
  {Miao}, \citenamefont {Ma}, \citenamefont {Richard}, \citenamefont {Huang},
  \citenamefont {Zhao}, \citenamefont {Chen}, \citenamefont {Fang},
  \citenamefont {Dai}, \citenamefont {Qian},\ and\ \citenamefont
  {Ding}}]{Weng:PhotCrystWSM}%
  \BibitemOpen
  \bibfield  {author} {\bibinfo {author} {\bibfnamefont {B.~Q.}\ \bibnamefont
  {Lv}}, \bibinfo {author} {\bibfnamefont {H.~M.}\ \bibnamefont {Weng}},
  \bibinfo {author} {\bibfnamefont {B.~B.}\ \bibnamefont {Fu}}, \bibinfo
  {author} {\bibfnamefont {X.~P.}\ \bibnamefont {Wang}}, \bibinfo {author}
  {\bibfnamefont {H.}~\bibnamefont {Miao}}, \bibinfo {author} {\bibfnamefont
  {J.}~\bibnamefont {Ma}}, \bibinfo {author} {\bibfnamefont {P.}~\bibnamefont
  {Richard}}, \bibinfo {author} {\bibfnamefont {X.~C.}\ \bibnamefont {Huang}},
  \bibinfo {author} {\bibfnamefont {L.~X.}\ \bibnamefont {Zhao}}, \bibinfo
  {author} {\bibfnamefont {G.~F.}\ \bibnamefont {Chen}}, \bibinfo {author}
  {\bibfnamefont {Z.}~\bibnamefont {Fang}}, \bibinfo {author} {\bibfnamefont
  {X.}~\bibnamefont {Dai}}, \bibinfo {author} {\bibfnamefont {T.}~\bibnamefont
  {Qian}}, \ and\ \bibinfo {author} {\bibfnamefont {H.}~\bibnamefont {Ding}},\
  }\href@noop {} {\bibfield  {journal} {\bibinfo  {journal} {Phys. Rev. X}\
  }\textbf {\bibinfo {volume} {5}},\ \bibinfo {pages} {031013} (\bibinfo {year}
  {2015})}\BibitemShut {NoStop}%
\bibitem [{\citenamefont {Borisenko}\ \emph {et~al.}(2014)\citenamefont
  {Borisenko}, \citenamefont {Gibson}, \citenamefont {Evtushinsky},
  \citenamefont {Zabolotnyy}, \citenamefont {B\"uchner},\ and\ \citenamefont
  {Cava}}]{Cava2014}%
  \BibitemOpen
  \bibfield  {author} {\bibinfo {author} {\bibfnamefont {S.}~\bibnamefont
  {Borisenko}}, \bibinfo {author} {\bibfnamefont {Q.}~\bibnamefont {Gibson}},
  \bibinfo {author} {\bibfnamefont {D.}~\bibnamefont {Evtushinsky}}, \bibinfo
  {author} {\bibfnamefont {V.}~\bibnamefont {Zabolotnyy}}, \bibinfo {author}
  {\bibfnamefont {B.}~\bibnamefont {B\"uchner}}, \ and\ \bibinfo {author}
  {\bibfnamefont {R.~J.}\ \bibnamefont {Cava}},\ }\href@noop {} {\bibfield
  {journal} {\bibinfo  {journal} {Phys. Rev. Lett.}\ }\textbf {\bibinfo
  {volume} {113}},\ \bibinfo {pages} {027603} (\bibinfo {year}
  {2014})}\BibitemShut {NoStop}%
\bibitem [{\citenamefont {Jeon}\ \emph {et~al.}(2014)\citenamefont {Jeon},
  \citenamefont {Zhou}, \citenamefont {Gyenis}, \citenamefont {Feldman},
  \citenamefont {Kimchi}, \citenamefont {Potter}, \citenamefont {Gibson},
  \citenamefont {Cava}, \citenamefont {Vishwanath},\ and\ \citenamefont
  {Yazdani}}]{Yazdani2014}%
  \BibitemOpen
  \bibfield  {author} {\bibinfo {author} {\bibfnamefont {S.}~\bibnamefont
  {Jeon}}, \bibinfo {author} {\bibfnamefont {B.~B.}\ \bibnamefont {Zhou}},
  \bibinfo {author} {\bibfnamefont {A.}~\bibnamefont {Gyenis}}, \bibinfo
  {author} {\bibfnamefont {B.~E.}\ \bibnamefont {Feldman}}, \bibinfo {author}
  {\bibfnamefont {I.}~\bibnamefont {Kimchi}}, \bibinfo {author} {\bibfnamefont
  {A.~C.}\ \bibnamefont {Potter}}, \bibinfo {author} {\bibfnamefont {Q.~D.}\
  \bibnamefont {Gibson}}, \bibinfo {author} {\bibfnamefont {R.~J.}\
  \bibnamefont {Cava}}, \bibinfo {author} {\bibfnamefont {A.}~\bibnamefont
  {Vishwanath}}, \ and\ \bibinfo {author} {\bibfnamefont {A.}~\bibnamefont
  {Yazdani}},\ }\href@noop {} {\bibfield  {journal} {\bibinfo  {journal} {Nat.
  Mat.}\ }\textbf {\bibinfo {volume} {13}},\ \bibinfo {pages} {851} (\bibinfo
  {year} {2014})}\BibitemShut {NoStop}%
\bibitem [{\citenamefont {Goswami}\ and\ \citenamefont
  {Chakravarty}(2011)}]{Goswami:TIRG}%
  \BibitemOpen
  \bibfield  {author} {\bibinfo {author} {\bibfnamefont {P.}~\bibnamefont
  {Goswami}}\ and\ \bibinfo {author} {\bibfnamefont {S.}~\bibnamefont
  {Chakravarty}},\ }\href@noop {} {\bibfield  {journal} {\bibinfo  {journal}
  {Phys. Rev. Lett.}\ }\textbf {\bibinfo {volume} {107}},\ \bibinfo {pages}
  {196803} (\bibinfo {year} {2011})}\BibitemShut {NoStop}%
\bibitem [{\citenamefont {Kobayashi}\ \emph {et~al.}(2014)\citenamefont
  {Kobayashi}, \citenamefont {Ohtsuki}, \citenamefont {Imura},\ and\
  \citenamefont {Herbut}}]{Herbut}%
  \BibitemOpen
  \bibfield  {author} {\bibinfo {author} {\bibfnamefont {K.}~\bibnamefont
  {Kobayashi}}, \bibinfo {author} {\bibfnamefont {T.}~\bibnamefont {Ohtsuki}},
  \bibinfo {author} {\bibfnamefont {K.-I.}\ \bibnamefont {Imura}}, \ and\
  \bibinfo {author} {\bibfnamefont {I.~F.}\ \bibnamefont {Herbut}},\
  }\href@noop {} {\bibfield  {journal} {\bibinfo  {journal} {Phys. Rev. Lett.}\
  }\textbf {\bibinfo {volume} {112}},\ \bibinfo {pages} {016402} (\bibinfo
  {year} {2014})}\BibitemShut {NoStop}%
\bibitem [{\citenamefont {Sbierski}\ \emph {et~al.}(2015)\citenamefont
  {Sbierski}, \citenamefont {Bergholtz},\ and\ \citenamefont
  {Brouwer}}]{Brouwer:exponents}%
  \BibitemOpen
  \bibfield  {author} {\bibinfo {author} {\bibfnamefont {B.}~\bibnamefont
  {Sbierski}}, \bibinfo {author} {\bibfnamefont {E.~J.}\ \bibnamefont
  {Bergholtz}}, \ and\ \bibinfo {author} {\bibfnamefont {P.~W.}\ \bibnamefont
  {Brouwer}},\ }\href@noop {} {\bibfield  {journal} {\bibinfo  {journal} {Phys.
  Rev. B}\ }\textbf {\bibinfo {volume} {92}},\ \bibinfo {pages} {115145}
  (\bibinfo {year} {2015})}\BibitemShut {NoStop}%
\bibitem [{\citenamefont {Sbierski}\ \emph {et~al.}(2014)\citenamefont
  {Sbierski}, \citenamefont {Pohl}, \citenamefont {Bergholtz},\ and\
  \citenamefont {Brouwer}}]{Brouwer:WSMcond}%
  \BibitemOpen
  \bibfield  {author} {\bibinfo {author} {\bibfnamefont {B.}~\bibnamefont
  {Sbierski}}, \bibinfo {author} {\bibfnamefont {G.}~\bibnamefont {Pohl}},
  \bibinfo {author} {\bibfnamefont {E.~J.}\ \bibnamefont {Bergholtz}}, \ and\
  \bibinfo {author} {\bibfnamefont {P.~W.}\ \bibnamefont {Brouwer}},\
  }\href@noop {} {\bibfield  {journal} {\bibinfo  {journal} {Phys. Rev. Lett.}\
  }\textbf {\bibinfo {volume} {113}},\ \bibinfo {pages} {026602} (\bibinfo
  {year} {2014})}\BibitemShut {NoStop}%
\bibitem [{\citenamefont {Roy}\ and\ \citenamefont {{Das
  Sarma}}(2014)}]{RoyDasSarma}%
  \BibitemOpen
  \bibfield  {author} {\bibinfo {author} {\bibfnamefont {B.}~\bibnamefont
  {Roy}}\ and\ \bibinfo {author} {\bibfnamefont {S.}~\bibnamefont {{Das
  Sarma}}},\ }\href@noop {} {\bibfield  {journal} {\bibinfo  {journal} {Phys.
  Rev. B}\ }\textbf {\bibinfo {volume} {90}},\ \bibinfo {pages} {241112(R)}
  (\bibinfo {year} {2014})}\BibitemShut {NoStop}%
\bibitem [{\citenamefont {Pixley}\ \emph
  {et~al.}(2016{\natexlab{a}})\citenamefont {Pixley}, \citenamefont {Goswami},\
  and\ \citenamefont {{Das Sarma}}}]{Pixley:ExactZ}%
  \BibitemOpen
  \bibfield  {author} {\bibinfo {author} {\bibfnamefont {J.~H.}\ \bibnamefont
  {Pixley}}, \bibinfo {author} {\bibfnamefont {P.}~\bibnamefont {Goswami}}, \
  and\ \bibinfo {author} {\bibfnamefont {S.}~\bibnamefont {{Das Sarma}}},\
  }\href@noop {} {\bibfield  {journal} {\bibinfo  {journal} {Phys. Rev. B}\
  }\textbf {\bibinfo {volume} {93}},\ \bibinfo {pages} {085103} (\bibinfo
  {year} {2016}{\natexlab{a}})}\BibitemShut {NoStop}%
\bibitem [{\citenamefont {Syzranov}\ \emph
  {et~al.}(2015{\natexlab{a}})\citenamefont {Syzranov}, \citenamefont
  {Radzihovsky},\ and\ \citenamefont {Gurarie}}]{Syzranov:Weyl}%
  \BibitemOpen
  \bibfield  {author} {\bibinfo {author} {\bibfnamefont {S.~V.}\ \bibnamefont
  {Syzranov}}, \bibinfo {author} {\bibfnamefont {L.}~\bibnamefont
  {Radzihovsky}}, \ and\ \bibinfo {author} {\bibfnamefont {V.}~\bibnamefont
  {Gurarie}},\ }\href@noop {} {\bibfield  {journal} {\bibinfo  {journal} {Phys.
  Rev. Lett.}\ }\textbf {\bibinfo {volume} {114}},\ \bibinfo {pages} {166601}
  (\bibinfo {year} {2015}{\natexlab{a}})}\BibitemShut {NoStop}%
\bibitem [{\citenamefont {Syzranov}\ \emph
  {et~al.}(2016{\natexlab{a}})\citenamefont {Syzranov}, \citenamefont
  {Ostrovsky}, \citenamefont {Gurarie},\ and\ \citenamefont
  {Radzihovsky}}]{Syzranov:exp}%
  \BibitemOpen
  \bibfield  {author} {\bibinfo {author} {\bibfnamefont {S.~V.}\ \bibnamefont
  {Syzranov}}, \bibinfo {author} {\bibfnamefont {P.~M.}\ \bibnamefont
  {Ostrovsky}}, \bibinfo {author} {\bibfnamefont {V.}~\bibnamefont {Gurarie}},
  \ and\ \bibinfo {author} {\bibfnamefont {L.}~\bibnamefont {Radzihovsky}},\
  }\href@noop {} {\bibfield  {journal} {\bibinfo  {journal} {Phys. Rev. B}\
  }\textbf {\bibinfo {volume} {93}},\ \bibinfo {pages} {155113} (\bibinfo
  {year} {2016}{\natexlab{a}})}\BibitemShut {NoStop}%
\bibitem [{\citenamefont {Syzranov}\ \emph
  {et~al.}(2015{\natexlab{b}})\citenamefont {Syzranov}, \citenamefont
  {Gurarie},\ and\ \citenamefont {Radzihovsky}}]{Syzranov:unconv}%
  \BibitemOpen
  \bibfield  {author} {\bibinfo {author} {\bibfnamefont {S.~V.}\ \bibnamefont
  {Syzranov}}, \bibinfo {author} {\bibfnamefont {V.}~\bibnamefont {Gurarie}}, \
  and\ \bibinfo {author} {\bibfnamefont {L.}~\bibnamefont {Radzihovsky}},\
  }\href@noop {} {\bibfield  {journal} {\bibinfo  {journal} {Phys. Rev. B}\
  }\textbf {\bibinfo {volume} {91}},\ \bibinfo {pages} {035133} (\bibinfo
  {year} {2015}{\natexlab{b}})}\BibitemShut {NoStop}%
\bibitem [{\citenamefont {G{\"a}rttner}\ \emph {et~al.}(2015)\citenamefont
  {G{\"a}rttner}, \citenamefont {Syzranov}, \citenamefont {Rey}, \citenamefont
  {Gurarie},\ and\ \citenamefont {Radzihovsky}}]{Garttner:longrange}%
  \BibitemOpen
  \bibfield  {author} {\bibinfo {author} {\bibfnamefont {M.}~\bibnamefont
  {G{\"a}rttner}}, \bibinfo {author} {\bibfnamefont {S.~V.}\ \bibnamefont
  {Syzranov}}, \bibinfo {author} {\bibfnamefont {A.~M.}\ \bibnamefont {Rey}},
  \bibinfo {author} {\bibfnamefont {V.}~\bibnamefont {Gurarie}}, \ and\
  \bibinfo {author} {\bibfnamefont {L.}~\bibnamefont {Radzihovsky}},\
  }\href@noop {} {\bibfield  {journal} {\bibinfo  {journal} {Phys. Rev. B}\
  }\textbf {\bibinfo {volume} {92}},\ \bibinfo {pages} {041406(R)} (\bibinfo
  {year} {2015})}\BibitemShut {NoStop}%
\bibitem [{\citenamefont {Ueoka}\ and\ \citenamefont
  {Slevin}(2014)}]{Slevin:high}%
  \BibitemOpen
  \bibfield  {author} {\bibinfo {author} {\bibfnamefont {Y.}~\bibnamefont
  {Ueoka}}\ and\ \bibinfo {author} {\bibfnamefont {K.}~\bibnamefont {Slevin}},\
  }\href@noop {} {\bibfield  {journal} {\bibinfo  {journal} {J. Phys. Soc.
  Japan}\ }\textbf {\bibinfo {volume} {83}},\ \bibinfo {pages} {084711}
  (\bibinfo {year} {2014})}\BibitemShut {NoStop}%
\bibitem [{\citenamefont {Syzranov}\ \emph
  {et~al.}(2016{\natexlab{b}})\citenamefont {Syzranov}, \citenamefont
  {Gurarie},\ and\ \citenamefont {Radzihovsky}}]{Syzranov:mult}%
  \BibitemOpen
  \bibfield  {author} {\bibinfo {author} {\bibfnamefont {S.~V.}\ \bibnamefont
  {Syzranov}}, \bibinfo {author} {\bibfnamefont {V.}~\bibnamefont {Gurarie}}, \
  and\ \bibinfo {author} {\bibfnamefont {L.}~\bibnamefont {Radzihovsky}},\
  }\href@noop {} {\bibfield  {journal} {\bibinfo  {journal} {Annals of
  Physics}\ }\textbf {\bibinfo {volume} {373}},\ \bibinfo {pages} {694}
  (\bibinfo {year} {2016}{\natexlab{b}})}\BibitemShut {NoStop}%
\bibitem [{\citenamefont {Foster}(2012)}]{Foster2012}%
  \BibitemOpen
  \bibfield  {author} {\bibinfo {author} {\bibfnamefont {M.}~\bibnamefont
  {Foster}},\ }\href@noop {} {\bibfield  {journal} {\bibinfo  {journal} {Phys.
  Rev. B}\ }\textbf {\bibinfo {volume} {85}},\ \bibinfo {pages} {085122}
  (\bibinfo {year} {2012})}\BibitemShut {NoStop}%
\bibitem [{\citenamefont {Nandkishore}\ \emph {et~al.}(2014)\citenamefont
  {Nandkishore}, \citenamefont {Huse},\ and\ \citenamefont
  {Sondhi}}]{Nandkishore:rare}%
  \BibitemOpen
  \bibfield  {author} {\bibinfo {author} {\bibfnamefont {R.}~\bibnamefont
  {Nandkishore}}, \bibinfo {author} {\bibfnamefont {D.~A.}\ \bibnamefont
  {Huse}}, \ and\ \bibinfo {author} {\bibfnamefont {S.~L.}\ \bibnamefont
  {Sondhi}},\ }\href@noop {} {\bibfield  {journal} {\bibinfo  {journal} {Phys.
  Rev. B}\ }\textbf {\bibinfo {volume} {89}},\ \bibinfo {pages} {245110}
  (\bibinfo {year} {2014})}\BibitemShut {NoStop}%
\bibitem [{\citenamefont {Pixley}\ \emph
  {et~al.}(2016{\natexlab{b}})\citenamefont {Pixley}, \citenamefont {Huse},\
  and\ \citenamefont {Sarma}}]{Pixley2016}%
  \BibitemOpen
  \bibfield  {author} {\bibinfo {author} {\bibfnamefont {J.~H.}\ \bibnamefont
  {Pixley}}, \bibinfo {author} {\bibfnamefont {D.~A.}\ \bibnamefont {Huse}}, \
  and\ \bibinfo {author} {\bibfnamefont {S.~D.}\ \bibnamefont {Sarma}},\
  }\href@noop {} {\bibfield  {journal} {\bibinfo  {journal} {Phys. Rev. X}\
  }\textbf {\bibinfo {volume} {6}},\ \bibinfo {pages} {021042} (\bibinfo {year}
  {2016}{\natexlab{b}})}\BibitemShut {NoStop}%
\bibitem [{\citenamefont {Pixley}\ \emph
  {et~al.}(2016{\natexlab{c}})\citenamefont {Pixley}, \citenamefont {Huse},\
  and\ \citenamefont {Das~Sarma}}]{Pixley2016a}%
  \BibitemOpen
  \bibfield  {author} {\bibinfo {author} {\bibfnamefont {J.~H.}\ \bibnamefont
  {Pixley}}, \bibinfo {author} {\bibfnamefont {D.~A.}\ \bibnamefont {Huse}}, \
  and\ \bibinfo {author} {\bibfnamefont {S.}~\bibnamefont {Das~Sarma}},\
  }\href@noop {} {\bibfield  {journal} {\bibinfo  {journal} {Phys. Rev. B}\
  }\textbf {\bibinfo {volume} {94}},\ \bibinfo {pages} {121107} (\bibinfo
  {year} {2016}{\natexlab{c}})}\BibitemShut {NoStop}%
\bibitem [{\citenamefont {Pixley}\ \emph {et~al.}(2017)\citenamefont {Pixley},
  \citenamefont {Chou}, \citenamefont {Goswami}, \citenamefont {Huse},
  \citenamefont {Nandkishore}, \citenamefont {Radzihovsky},\ and\ \citenamefont
  {{Das Sarma}}}]{Radzihovsky2017}%
  \BibitemOpen
  \bibfield  {author} {\bibinfo {author} {\bibfnamefont {J.~H.}\ \bibnamefont
  {Pixley}}, \bibinfo {author} {\bibfnamefont {Y.-Z.}\ \bibnamefont {Chou}},
  \bibinfo {author} {\bibfnamefont {P.}~\bibnamefont {Goswami}}, \bibinfo
  {author} {\bibfnamefont {D.~A.}\ \bibnamefont {Huse}}, \bibinfo {author}
  {\bibfnamefont {R.}~\bibnamefont {Nandkishore}}, \bibinfo {author}
  {\bibfnamefont {L.}~\bibnamefont {Radzihovsky}}, \ and\ \bibinfo {author}
  {\bibfnamefont {S.}~\bibnamefont {{Das Sarma}}},\ }\href@noop {} {\enquote
  {\bibinfo {title} {Single particle excitations in disordered {Weyl}
  fluids},}\ } (\bibinfo {year} {2017}),\ \bibinfo {note}
  {arXiv:1701.00783}\BibitemShut {NoStop}%
\bibitem [{\citenamefont {Holder}\ \emph {et~al.}(2017)\citenamefont {Holder},
  \citenamefont {Huang},\ and\ \citenamefont {Ostrovsky}}]{Ostrovsky2017}%
  \BibitemOpen
  \bibfield  {author} {\bibinfo {author} {\bibfnamefont {T.}~\bibnamefont
  {Holder}}, \bibinfo {author} {\bibfnamefont {C.-W.}\ \bibnamefont {Huang}}, \
  and\ \bibinfo {author} {\bibfnamefont {P.~M.}\ \bibnamefont {Ostrovsky}},\
  }\href@noop {} {\enquote {\bibinfo {title} {Electronic properties of
  disordered {Weyl} semimetals at charge neutrality},}\ } (\bibinfo {year}
  {2017}),\ \bibinfo {note} {arXiv:1704.05481}\BibitemShut {NoStop}%
\bibitem [{Note1()}]{Note1}%
  \BibitemOpen
  \bibinfo {note} {The correlation length, being the length of the spatial
  decay of the single particle Green's function, should be identified with the
  mean free path of a particle in disordered environment. From this point of
  view, it is indeed natural that it does not diverge to infinity; otherwise
  one would be forced to conclude that the mean free path could be infinite at
  a finite disorder strength.}\BibitemShut {Stop}%
\bibitem [{\citenamefont {Gennes}(1972)}]{deGennes1972}%
  \BibitemOpen
  \bibfield  {author} {\bibinfo {author} {\bibfnamefont {P.~G.~D.}\
  \bibnamefont {Gennes}},\ }\href@noop {} {\bibfield  {journal} {\bibinfo
  {journal} {Phys. Lett. A}\ }\textbf {\bibinfo {volume} {38}},\ \bibinfo
  {pages} {339} (\bibinfo {year} {1972})}\BibitemShut {NoStop}%
\bibitem [{\citenamefont {Emery}(1975)}]{Emery1975}%
  \BibitemOpen
  \bibfield  {author} {\bibinfo {author} {\bibfnamefont {V.~J.}\ \bibnamefont
  {Emery}},\ }\href@noop {} {\bibfield  {journal} {\bibinfo  {journal} {Phys.
  Rev. B}\ }\textbf {\bibinfo {volume} {11}},\ \bibinfo {pages} {239} (\bibinfo
  {year} {1975})}\BibitemShut {NoStop}%
\bibitem [{\citenamefont {Oono}(2075)}]{Oono1975}%
  \BibitemOpen
  \bibfield  {author} {\bibinfo {author} {\bibfnamefont {Y.}~\bibnamefont
  {Oono}},\ }\href@noop {} {\bibfield  {journal} {\bibinfo  {journal} {J. Phys.
  Soc. Japan}\ }\textbf {\bibinfo {volume} {39}},\ \bibinfo {pages} {25}
  (\bibinfo {year} {2075})}\BibitemShut {NoStop}%
\bibitem [{\citenamefont {Oono}(1976)}]{Oono1976}%
  \BibitemOpen
  \bibfield  {author} {\bibinfo {author} {\bibfnamefont {Y.}~\bibnamefont
  {Oono}},\ }\href@noop {} {\bibfield  {journal} {\bibinfo  {journal} {J. Phys.
  Soc. Japan}\ }\textbf {\bibinfo {volume} {41}},\ \bibinfo {pages} {787}
  (\bibinfo {year} {1976})}\BibitemShut {NoStop}%
\bibitem [{\citenamefont {Brydges}\ and\ \citenamefont
  {Slade}(1995)}]{Slade1995}%
  \BibitemOpen
  \bibfield  {author} {\bibinfo {author} {\bibfnamefont {D.~C.}\ \bibnamefont
  {Brydges}}\ and\ \bibinfo {author} {\bibfnamefont {G.}~\bibnamefont
  {Slade}},\ }\href@noop {} {\bibfield  {journal} {\bibinfo  {journal} {Prob.
  Theory Relat. Fields}\ }\textbf {\bibinfo {volume} {103}},\ \bibinfo {pages}
  {285} (\bibinfo {year} {1995})}\BibitemShut {NoStop}%
\bibitem [{\citenamefont {Lifshitz}(1963)}]{Lifshitz:tail}%
  \BibitemOpen
  \bibfield  {author} {\bibinfo {author} {\bibfnamefont {I.~M.}\ \bibnamefont
  {Lifshitz}},\ }\href@noop {} {\bibfield  {journal} {\bibinfo  {journal} {Sov.
  Phys. JETP}\ }\textbf {\bibinfo {volume} {17}},\ \bibinfo {pages} {1159}
  (\bibinfo {year} {1963})}\BibitemShut {NoStop}%
\bibitem [{\citenamefont {Zittartz}\ and\ \citenamefont
  {Langer}(1966)}]{ZittartzLanger}%
  \BibitemOpen
  \bibfield  {author} {\bibinfo {author} {\bibfnamefont {J.}~\bibnamefont
  {Zittartz}}\ and\ \bibinfo {author} {\bibfnamefont {J.~S.}\ \bibnamefont
  {Langer}},\ }\href@noop {} {\bibfield  {journal} {\bibinfo  {journal} {Phys.
  Rev.}\ }\textbf {\bibinfo {volume} {148}},\ \bibinfo {pages} {741} (\bibinfo
  {year} {1966})}\BibitemShut {NoStop}%
\bibitem [{\citenamefont {Cardy}(1978)}]{Cardy1978}%
  \BibitemOpen
  \bibfield  {author} {\bibinfo {author} {\bibfnamefont {J.}~\bibnamefont
  {Cardy}},\ }\href@noop {} {\bibfield  {journal} {\bibinfo  {journal} {J.
  Phys. C Solid State}\ }\textbf {\bibinfo {volume} {11}} (\bibinfo {year}
  {1978})}\BibitemShut {NoStop}%
\bibitem [{\citenamefont {Coleman}\ \emph {et~al.}(1978)\citenamefont
  {Coleman}, \citenamefont {Glaser},\ and\ \citenamefont
  {Martin}}]{Coleman1978}%
  \BibitemOpen
  \bibfield  {author} {\bibinfo {author} {\bibfnamefont {S.}~\bibnamefont
  {Coleman}}, \bibinfo {author} {\bibfnamefont {V.}~\bibnamefont {Glaser}}, \
  and\ \bibinfo {author} {\bibfnamefont {A.}~\bibnamefont {Martin}},\
  }\href@noop {} {\bibfield  {journal} {\bibinfo  {journal} {Comm. Math.
  Phys.}\ }\textbf {\bibinfo {volume} {58}},\ \bibinfo {pages} {211} (\bibinfo
  {year} {1978})}\BibitemShut {NoStop}%
\bibitem [{\citenamefont {Suslov}(1994)}]{Suslov1994}%
  \BibitemOpen
  \bibfield  {author} {\bibinfo {author} {\bibfnamefont {I.~M.}\ \bibnamefont
  {Suslov}},\ }\href@noop {} {\bibfield  {journal} {\bibinfo  {journal} {Zh.
  Eksp. Teor. Fiz.}\ }\textbf {\bibinfo {volume} {106}},\ \bibinfo {pages}
  {560} (\bibinfo {year} {1994})}\BibitemShut {NoStop}%
\bibitem [{\citenamefont {Polyakov}(1977)}]{Polyakov1977}%
  \BibitemOpen
  \bibfield  {author} {\bibinfo {author} {\bibfnamefont {A.}~\bibnamefont
  {Polyakov}},\ }\href@noop {} {\bibfield  {journal} {\bibinfo  {journal}
  {Nucl. Phys. B}\ }\textbf {\bibinfo {volume} {120}},\ \bibinfo {pages} {429}
  (\bibinfo {year} {1977})}\BibitemShut {NoStop}%
\bibitem [{\citenamefont {Polyakov}(1987)}]{Polyakov:book}%
  \BibitemOpen
  \bibfield  {author} {\bibinfo {author} {\bibfnamefont {A.~M.}\ \bibnamefont
  {Polyakov}},\ }\href@noop {} {\emph {\bibinfo {title} {Gauge Fields and
  Strings}}}\ (\bibinfo  {publisher} {CRC Press},\ \bibinfo {address} {New
  York},\ \bibinfo {year} {1987})\BibitemShut {NoStop}%
\end{thebibliography}%




\onecolumngrid

\appendix
\section{Coulomb gas action}
\label{sec:A}
We would like to elaborate on how  to derive the effective action of the Coulomb gas for the disordered Schr\"odinger equation defined by Eqs.~\rf{eq:schr} and \rf{eq:dc} in $d>d_c=4$ dimensions. Almost all the steps in this derivation are standard, although their implementation in this problem have some peculiarities which will be emphasized here.

The energy of a gas of instantons can be found via substituting \rfs{eq:gas} into \rfs{eq:actionkernel}. 
\be  \label{eq:actionexplained} S = \frac{\lambda}{4} \int \frac{d^d r d^d r'}{\xi^d} \left( \sum_{n=1}^N \zeta_n f_{\rm inst} \left( {\bf r}-{\bf r}_n \right) \right)^2  \left( \sum_{m=1}^N \zeta_m f_{\rm inst} \left( {\bf r}'-{\bf r}_m \right) \right)^2 K\left(\frac{ \left| {\bf r}-{\bf r}' \right| }{\xi} \right),
\ee
where $\zeta_n=\pm 1$, so that $\zeta_n^2=1$. 

Opening the brackets, the first term is
\be S_0= \frac{\lambda}{4} \sum_{n=1}^N \int \frac{d^d r d^d r'}{\xi^d}  f^2_{\rm inst} \left( {\bf r}-{\bf r}_n \right)  f^2_{\rm inst} \left( {\bf r}'-{\bf r}_n \right)  K\left(\frac{ \left| {\bf r}-{\bf r}' \right| }{\xi} \right) =
C_1 N \frac{\xi^{d-4}}{\lambda}.
\ee
To do this calculation we replaced $f_{\rm inst}$ by its asymptotic behavior \rfs{eq:finst}, replaced the kernel $K$ by a delta function and estimated the integral, up to an overall numerical 
coefficient $C_1$, by integrating up to distances $\xi$. 
 This is nothing but the sum of the ``core energies" of instantons \rfs{eq:singleact}, one per instanton. The next term has the form
 \be  S_1\sim \lambda \sum_{n \not = m} \zeta_n \zeta_m \frac{\int d^d r d^d r'}{\xi^d}   f^2_{\rm inst} \left( {\bf r}-{\bf r}_n \right)  f_{\rm inst} \left( {\bf r}'-{\bf r}_n \right)  f_{\rm inst} \left( {\bf r}'-{\bf r}_m \right) K\left(\frac{ \left| {\bf r}-{\bf r}' \right| }{\xi} \right). \ee
 In its estimation we again replace the kernel $K$ by the delta-function and use the asymptotic behavior \rfs{eq:finst}, to find
 \be \label{eq:twob} S_1 \approx A_0^4 \frac {\xi^{4d-12}}{\lambda} \sum_{n \not = m} \zeta_n \zeta_m \int \frac{d^d r}{\left| {\bf r}- {\bf r}_n \right|^{3d-6} \left| {\bf r}- {\bf r}_m \right|^{d-2}},
\ee
where the integral is computed over the domain excluding the circles of radius $\xi$ surrounding each of the points ${\bf r}_n$, ${\bf r}_m$. 
 We observe that this integral is then dominated by ${\bf r}$ in the vicinity
 of ${\bf r}_n$ as long as $d>3$ which is definitely the case here, where we find
\be \label{eq:twobo} S_1 =C_2 \frac{\xi^{2d -6}}{\lambda} \sum_{n \not =  m}   \frac{\zeta_n \zeta_m}{\left| {\bf r}_n- {\bf r}_m \right|^{d-2}}.
\ee
as stated in \rfs{eq:instint}. Importantly, as ${\bf r}$ approaches ${\bf r}_m$ in \rfs{eq:twob} the integral there is not divergent and so the contribution of the neighborhood of ${\bf r}_m$ to $S_1$ will be small and can be neglected at small $\xi$. Note that this expression is correct only if $\left| {\bf r}_n- {\bf r}_m \right| \gg \xi$. 

Furthermore, all other terms arising from \rfs{eq:actionexplained} have weaker divergencies than the ones leading to \rfs{eq:twobo} and so can be neglected. Thus we obtained a Coulomb gas
of charges, which can be treated using the standard methods. These methods involve the following steps. We take advantage of the Gaussian integral representation of the Coulomb gas
\be \int {\cal D} \chi \, e^{- \oh \lambda \xi^{6-2d} C_4 \int d^d x \left( \nabla \chi \right)^2  + i \sum_n \zeta_n \chi({\bf r}_n) } =\exp \left( -C_2 \frac{\xi^{2d-6}}{\lambda} \sum_{n \not =  m} \frac{\zeta_n \zeta_m}{\left| {\bf r}_n- {\bf r}_m \right|^{d-2}} -N C_3 \frac{\xi^{d-4}}{\lambda} \right)
\ee
The coefficient $C_4$ is chosen to produce $C_2$ on the right hand side, which is straightforward to do explicitly but is not necessary here. The numerical coefficient $C_3$ arises as the term $\chi({\bf r}_n)$ in the exponential on the left hand side should be understood as $\chi({\bf r})$ integrated together with an envelope function centered around ${\bf r}_n$ designed to
produce $S_1$ not only when $\left| {\bf r}_n -{\bf r}_m \right| \gg \xi$, but also when these two points approach each other. This regularizes $1/\left| {\bf r}_n- {\bf r}_m\right|^{d-2}$ at distances less than $\xi$
as roughly $1/\xi^{d-2}$, up to the coefficient $C_3$. 

This allows us to sum over sectors with arbitrary number of instantons, and sum over $\zeta_n=\pm$, in the standard way with the result
\be \label{eq:effe} \int {\cal D} \chi \, e^{-   \int d^d r \left[\frac{D \lambda \xi^{6-2d}}{2}  \left( \nabla \chi \right)^2 - i \mu \xi^{-d} \cos\left(\chi \right)  \right]}.
\ee
Here 
\be \mu = C_5 e^{-C_4 \frac{\xi^{d-4}}{\lambda}}.
\ee
Here the coefficient $C_4=C_1-C_3$. As emphasized in Ref.~[\onlinecite{Cardy1978}], $i \mu$ is purely imaginary. This point is subtle: this arises from integrating over Gaussian fluctuations about the
instanton solution in the limit of number of replicas taken to zero. In Ref.~[\onlinecite{Cardy1978}] this produced an overall coefficient $i$. In our multi-instanton calculation this produces a factor of $i^N$ 
in the sector with $N$ instantons, reproduced by expanding \rfs{eq:effe} in powers of $\mu$. 

In the theory given by \rfs{eq:effe} $\cos(\chi)$ is always a relevant perturbation as long as $d\ge 3$ (unlike the conventional BKT transition where it can be either relevant or irrelevant). Since it is relevant,
we can expand the exponential in \rfs{eq:effe} in powers of $\chi$ to find the effective theory given by
\be \int {\cal D} \chi \, e^{- \oh   \int d^d r \left[D \lambda \xi^{6-2d} \left( \nabla \chi \right)^2 + i \mu \xi^{-d}   \chi^2  \right]}.
\ee
In turn, this produces the Green's function
\be \label{eq:grrr} G(p)  \sim \frac{1}{D \lambda \xi^{6-2d} p^2 +i \mu \xi^{-d}}
\ee
with purely imaginary $\mu$. While intuitively this expression should be correct, one can follow the prescription worked out in Ref.~[\onlinecite{Polyakov:book}] to resum over
the instanton contribution to arrive at \rfs{eq:grrr} as the Green's function of the field $\phi_1$ of \rfs{eq:part}. This arguments proceeds in the following way: we express the field $\phi$ from \rfs{eq:part}
as
\be \label{eq:dec} \phi_n({\bf R})  =  v_n \sqrt{-i} \sum_{m=1}^N f_{\rm inst} \left( {\bf R}- {\bf r}_m \right) \zeta_m + \varphi_n ({\bf R})  =  v_n \sqrt{-1}\int d^3 r f_{\rm inst} \left({\bf R} - {\bf  r} \right) \rho_{\rm inst}({\bf r})
+ \varphi_n ({\bf R}).
\ee
Here $N$ is the number of instantons, $n$ is the replica index, and $\varphi_n$ are the Gaussian fluctuations about the instanton solution. $\varphi$ gives perturbative contributions to the correlator of $\phi$.  $\rho_{\rm inst}$ is the instanton density defined by
\be \rho_{\rm inst} = \sum_{m=1}^N \delta \left({\bf r}-{\bf r}_m \right) \zeta_m.
\ee
Now the correlations of $\rho_{\rm inst}$ can be extracted from the generating functional \cite{Polyakov:book}
\be \left< e^{i  \int d^3 r \rho_{\rm inst}({\bf r}) \eta({\bf r})} \right> = \int {\cal D} \chi \, e^{-  \oh  \int d^d r \left[D \lambda \xi^{6-2d}  \left( \nabla (\chi-\eta) \right)^2 + i \mu \xi^{-d} \cos\left(\chi \right)  \right]}.
\ee
It follows that (by varying with respect to $\eta$)
$$ \left< \rho_{\rm inst}(p) \rho_{\rm inst}(-p) \right>=-D \lambda \xi^{6-2d} p^2 + \left( D \lambda \xi^{6-2d} p^2 \right)^2 \left< \chi(p) \chi(-p) \right> =-D \lambda \xi^{6-2d} p^2 + \left( D \lambda \xi^{6-2d} p^2 \right)^2  \frac{1} {D \lambda \xi^{6-2d} p^2 + i  \mu \xi^{-d}}=
$$
\be = -\frac{i \mu D \lambda \xi^{6-3d} p^2 }{D \lambda \xi^{6-2d} p^2 +i \mu \xi^{-d}}.
\ee
In momentum space, \rfs{eq:dec} reads
\be \phi_n = v_n \sqrt{i} \frac{B \xi^{d-3}}{\sqrt{\lambda} p^2} \rho_{\rm inst} ({\bf p}),
\ee
where $B$ is yet another dimensionless factor related to the Fourier transform of \rfs{eq:finst}. This leads to the contribution of the instantons to the Green's function
\be \label{eq:sum} G_{\rm inst}= \frac{1}{p^2} \frac{ i \mu \xi^{-d} D B^2 }{D \lambda \xi^{6-2d} p^2 + i \mu \xi^{-d}}.
\ee
We should add to this the contribution of the instanton-free sector, which his $-1/p^2$. We expect that this eliminates the pole at $p\rightarrow 0$ in \rfs{eq:sum} or that $DB^2=1$, to give
\be \label{eq:greenfull} G =  \frac{1}{p^2} \left( \frac{ i \mu  \xi^{-d}D B^2 }{D \lambda \xi^{6-2d} p^2 + i \mu \xi^{-d} }- 1 \right) = -\frac{D \lambda \xi^{6-2d} }{D \lambda \xi^{6-2d} p^2 + i \mu \xi^{-d}},
\ee
the result quoted above in \rfs{eq:grrr} and in \rfs{eq:gmain}.

This produces the correlation length
\be \label{eq:corlength} \ell \sim \frac{\xi^{3-d/2} \sqrt{\lambda}}{\sqrt{\mu}}.
\ee
This correlation length results in  the avoidance of criticality in the random Schr\"odinger problem, with the critical scaling occurring below this length only.

Note that this same Green's function can be used to evaluate the density of states. At $d>d_c=4$, we find
\be \label{eq:dee} \rho(0) = \frac  1 \pi {\rm Im} \, \int \frac{d^d p}{(2\pi)^d}  \frac{D \lambda \xi^{6-2d} }{D \lambda \xi^{6-2d} p^2 + i \mu \xi^{-d}} \approx \frac{\mu}{D\lambda \xi^{6-d} \pi }  \int \frac{d^d p}{(2\pi)^d} \frac{1}{p^4} \sim \frac{\mu}{\lambda \xi^{2}}.
\ee
Here we use that the integral on the right hand side of \rf{eq:dee} is divergent at large $p$ and should be cut off at $p \sim 1/\xi$. Comparison between Eqs.~\rf{eq:corlength}, \rf{eq:dee} gives
\be \ell \sim \frac {\xi^{2-\frac d 2}} {\sqrt{\rho(0)} },
\ee  the relationship between the density of states at zero energy (where it would've been zero in the absence of disorder) and the correlation length built into the Green's function. This relationship conveniently relates the scale at which criticality is avoided with the density of states produced by the rare fluctuations of disorder. 

\section{Instanton gas for arbitrary $\alpha$}
\label{sec:B}

We would like to discuss briefly the generalization of the above formalism to the case when the spectrum in the absence of disorder is $E=p^\alpha$. In particular, $\alpha=2$ for the Schr\"odinger problem, $\alpha=1$ for the Weyl/Dirac problem, and in other examples $\alpha$ could take arbitrary values. In order for weak disorder to be irrelevant and in order for the RG critical point to exist, $d>d_c=2\alpha$. 

A single instanton in this problem goes as $f_{\rm inst} \sim 1/r^{d-\alpha}$. This can be seen from the fact that at distances $r \gg \xi$ the instanton coincides with the Green's function of the kinetic energy operator \cite{Nandkishore:rare}, which is $1/p^\alpha$ in momentum space. Marching the cubic and linear terms in the corresponding generalization of \rfs{eq:nsi}, we find that
\be f_{\rm inst} = A_0 \frac{\xi^{d-3\alpha/2}}{r^{d-\alpha} \sqrt{\lambda}}, \ r \gg \xi.
\ee
This is the generalization of \rfs{eq:finst}. Evaluating the action of the single instanton gives a generalization of \rfs{eq:singleact} or $S \sim \xi^{d-2\alpha}/\lambda$ leading to the density of states
\be \rho \sim \exp \left( - {\rm const}  \, \xi^{d-2\alpha}/\lambda \right).
\ee
A gas of such instantons can now be constructed, which interacts via a pair potential $\xi^{2d-3\alpha}/(\lambda r^{d-\alpha})$, generalizing \rfs{eq:twobo}. In order to derive this, we need a divergence
of the corresponding integral in \rfs{eq:twob}, which requires $d>3\alpha/2$, which is indeed fulfilled here due to $d>2\alpha$. This produces an effective field theory 
\be \int {\cal D} \chi \, e^{- \oh  \int d^d r \left[ D \lambda \xi^{3 \alpha -2d}  \chi \left| \d \right|^{\alpha} \chi + i \mu \xi^{-d}  \chi^2  \right]},
\ee
with 
\be \mu = C_5 e^{-C_4 \frac{\xi^{d-2\alpha}}{\lambda}}.
\ee
By analogy with this gives the Green's function \rfs{eq:gralpj}, given also here for reference
\be \label{eq:gralpj1} G(p) \sim\frac{ D \lambda \xi^{3 \alpha-2d}}{D \lambda \xi^{3 \alpha-2d} p^\alpha + i \mu \xi^{-d} },
\ee

Finally, the correlation length in this theory is 
\be \ell \sim \xi^{3-d/\alpha} \left( \frac{\lambda}{\mu} \right)^{1/\alpha}.
\ee
The density of states is 
calculated in the same way as previously, 
\be \rho(0) \sim {\rm Im} \int \frac{d^d p}{(2\pi)^d} \frac{1}{p^\alpha + i \mu \xi^{d-3 \alpha}/(D \lambda)} \sim \frac{\mu}{\lambda} \xi^{d-3 \alpha} \int \frac{d^d p}{p^{2 \alpha}} \sim \frac{\mu}{\xi^{\alpha} \lambda}. 
\ee
Just as was done earlier, the integral $d^dp/p^{2 \alpha}$ can be estimated by the substitution of its upper limit where $p =1/\xi$, since $d>2\alpha$. 
This is still proportional to $\mu$, resulting in the relation
\be \ell \sim \frac 1 { \xi^{\frac d \alpha - 2} \rho(0)^{1/\alpha}},
\ee
given in the main text of this paper as \rfs{eq:dosalll}.
In particular, in the Weyl problem \rfs{eq:Weyl} the correlation length at what would have been a critical point goes as the inverse density of states induced by the rare fluctuations of disorder at the Dirac point.

\end{document}